\documentclass[conference]{IEEEtran}
\IEEEoverridecommandlockouts

\usepackage[utf8]{inputenc}
\usepackage[T1]{fontenc}
\usepackage{amsmath,amssymb,amsfonts}
\usepackage{graphicx}
\usepackage{textcomp}
\usepackage{xcolor}
\usepackage{hyperref}
\usepackage{booktabs}
\usepackage[numbers,sort&compress]{natbib}
\usepackage{tabularx}
\usepackage{booktabs}
\usepackage{array}
\usepackage{ragged2e}
\usepackage{enumitem}

\newcolumntype{Y}{>{\RaggedRight\arraybackslash}X}

\def\BibTeX{{\rm B\kern-.05em{\sc i\kern-.025em b}\kern-.08em
    \sc i\kern-.025em b\kern-.08em \sc i\kern-.025em b\kern-.05em}} 

\begin{document}

\title{ADEMM: A Longitudinal Method for Monitoring Developer Efficiency in Industry}

\author{
  \IEEEauthorblockN{Danilo Ribeiro\IEEEauthorrefmark{1}\IEEEauthorrefmark{2},
                    Breno Alves\IEEEauthorrefmark{1},
                    Gabriel
                    Souza
                    \IEEEauthorrefmark{1},
                    César França\IEEEauthorrefmark{1},
                    Alberto Souza\IEEEauthorrefmark{3}}
  \IEEEauthorblockA{\IEEEauthorrefmark{1}Franssa\\
  Recife, Brazil\\
  \{danilo, breno, gabriel, cesar\}@franssa.com}
  \IEEEauthorblockA{\IEEEauthorrefmark{2}CESAR School\\
  Recife, Brazil}
  \IEEEauthorblockA{\IEEEauthorrefmark{3}DevEficiente\\
  São Paulo, São Paulo\\
  alberto@deveficiente.com}
}

\maketitle

\begin{abstract}

\textbf{Context:} Developer efficiency is influenced by technical, organizational, cognitive, and communication-related factors. However, most studies rely on one-time assessments or fixed instruments, limiting the ability to monitor how barriers emerge and change over time, especially in consulting and professional education contexts.

\textbf{Objective:} This study proposes and evaluates the Adaptive Developer Efficiency Monitoring Method (ADEMM), an adaptive longitudinal method for monitoring developer efficiency when the monitoring organization does not directly employ the developers.

\textbf{Method:} Following Design Science Research and Action Design Research, we conducted a mixed-method longitudinal study with 27 software developers over twelve survey cycles. ADEMM was designed and refined through five iterative cycles, combining recurring surveys, 18 semi-structured interviews, and joint evaluation with a problem owner.

\textbf{Results:} The study resulted in ADEMM, a method that supports continuous data collection, mixed-methods integration, and iterative redesign of monitoring instruments. The evaluation produced three design principles: prioritization with the problem owner based on actionability, combination of closed and open data collection, and adaptation of items based on low variance and emerging qualitative signals.

\textbf{Conclusions:} ADEMM provides a transferable approach for adaptive longitudinal monitoring of developer efficiency. It helps balance comparability, contextual sensitivity, and practical utility in environments where organizations need to support developers without directly controlling their work contexts.

\end{abstract}

\begin{IEEEkeywords}
developer efficiency, longitudinal study, adaptive instrument design,
empirical software engineering, perceived barriers, generative AI
\end{IEEEkeywords}


\section{Introduction}
\label{sec:introduction}


Tracking software developers' efficiency over time is a recognized practical
need, but one that remains methodologically underresolved.
The literature establishes that software development efficiency cannot be
reduced to a single metric or dimension~\cite{forsgren2021space,sadowski2019rethinking}.
Efficiency results from the interaction of technical factors, such as
infrastructure quality and task complexity; organizational factors, such as
priority clarity, cross-team dependencies, and communication flow; and
cognitive factors, such as mental load, psychological safety, and flow
state~\cite{greiler2023devex,demarco1987peopleware,forsgren2018accelerate}.
Organizations that monitor these factors continuously are able to identify
specific barriers, allocate interventions more precisely, and evaluate the
effect of those interventions over time~\cite{forsgren2021space}.

This scenario has become more complex with the rapid adoption of Generative
Artificial Intelligence (GenAI) tools in developers' daily
work~\cite{kalliamvakou2022}.
These tools introduced new categories of friction, such as difficulty
obtaining adequate responses from models and time spent validating
automatically generated code, which were not present in monitoring
instruments developed before their widespread adoption.
The phenomenon to be monitored is therefore not static: the barriers to
efficiency change in response to technological, organizational, and
contextual shifts, which imposes specific requirements on the data
collection method used to assess developer efficiency.


Research on developer efficiency has produced significant advances in recent
years, but it remains limited in three interrelated respects.

The first concerns the nature of the instruments used.
Most empirical studies employ collection instruments defined entirely before
the investigation begins and kept fixed throughout the observation
period~\cite{meyer2019,russo2024longitudinal,kuutila2021individual}.
This choice is reasonable when the phenomenon is stable, but it becomes
problematic when new dimensions emerge during the investigation itself.
In the study by Kuutila et al.~\cite{kuutila2021individual}, for example,
the \textit{experience sampling} instrument was defined a priori and kept
constant throughout eight months of data collection; the same occurs in the
studies by Russo et al.~\cite{russo2024longitudinal} and
Meyer et al.~\cite{meyer2019}.
None of these works describes an explicit mechanism for incorporating into
the instrument factors that the qualitative data themselves reveal as
relevant during collection, which means that emerging dimensions
systematically fall outside the scope of observation until a new study
begins.

The second aspect concerns access to the work environment.
The most robust longitudinal studies identified in the literature combine
survey data with metrics extracted directly from repositories, continuous
integration systems, or internal tool logs of the investigated
organizations~\cite{kuutila2021individual,meyer2019}.
This combination increases the validity of the measures, but requires the
researcher to belong to the organization or to have formal access to its
infrastructure.
The problem is that this additional validity depends on a structural
condition, employment ties or access to internal infrastructure, that
consulting firms, professional education providers, and continuing training
programs do not have by the very definition of their business model: these
organizations follow developers without directly employing them.
In these contexts, the monitoring instrument needs to function
independently, relying exclusively on the developers' own perceptions of
their work.
The literature offers few models for this configuration, and the ones that
exist do not describe how the instrument should evolve over
time~\cite{canedo2019factors}.

The third aspect concerns transferability.
The longitudinal studies identified describe specific investigations
conducted in particular contexts, but do not formalize the monitoring
process as a replicable method.
As a result, an organization wishing to conduct a similar follow-up has no
set of steps, instrument review criteria, or design principles it can
instantiate in its own context.
The absence of this level of formalization limits the community's ability
to accumulate knowledge about what works and what does not work in the
longitudinal monitoring of developer efficiency.


To address these gaps, this paper proposes and evaluates the
\textit{Adaptive Developer Efficiency Monitoring Method} (ADEMM),
a longitudinal monitoring method designed to continuously track the factors
that affect developer efficiency in contexts in which the organization
responsible for the monitoring does not employ the professionals being
monitored, nor has direct access to their work environment.
ADEMM is grounded in the principles of \textit{Design Science Research}
(DSR)~\cite{peffers2007design,hevner2004design} and \textit{Action Design
Research} (ADR)~\cite{Sein2011} to structure an iterative process of
building, evaluating, and refining the monitoring instrument.
Unlike fixed instruments, ADEMM defines explicit criteria for removing items
with decreasing variability and for incorporating new dimensions identified
during the investigation, allowing the instrument to track the observed
phenomenon without requiring access to internal data from the employing
organizations.

The research question guiding this paper is:

\begin{itemize}
    \item \textbf{RQ:} How can developer efficiency be monitored longitudinally and adaptively without direct access to the work environment?
\end{itemize}


This paper offers four contributions to Empirical Software Engineering.
First, it proposes ADEMM as an adaptive longitudinal monitoring method,
describing its steps, instrument evolution criteria, and applicability
conditions in consulting and professional education contexts.
Second, it formalizes three design principles derived from the iterative
evolution of the method over twelve collection cycles, with 27 developers
working at distinct organizations.
Third, it demonstrates how explicit instrument review criteria, based on
mixed quantitative and qualitative evidence, enable the incorporation of
emerging dimensions during the investigation itself.
Fourth, it offers a documented instantiation of the method that serves as a
reference for researchers and organizations interested in conducting similar
monitoring efforts in similar contexts.

The remainder of the paper is organized as follows.
Section~\ref{sec:background} presents the theoretical background on
developer efficiency, longitudinal monitoring, DSR, and ADR.
Section~\ref{sec:method} describes the research design and the application
context of the method.
Section~\ref{sec:ademm-artifact} presents ADEMM, detailing its components,
review criteria, and instrument evolution cycles.
Section~\ref{sec:results} presents the formalized design principles.
Section~\ref{sec:discussion} discusses the method's transferability and its
implications for research and practice.
Section~\ref{sec:threats-validity} presents the threats to validity.
Section~\ref{sec:conclusion} concludes the paper and indicates future
directions.

\section{Background}
\label{sec:background}

\subsection{Developer Efficiency}

Developer productivity and efficiency are related but distinct concepts. While productivity refers to the outcomes achieved through software development work, efficiency emphasizes the relationship between those outcomes and the time, effort, and resources required to achieve them \cite{meyer2014software,coelho2025software}. In practical terms, efficiency means producing high-quality work with minimal friction and unnecessary effort. Consequently, completing more tasks does not necessarily indicate higher efficiency when developers must contend with excessive waiting, rework, interruptions, or unnecessary cognitive effort \cite{noda2023devex,coelho2025software}.

This distinction has a direct anchor in the developer productivity
measurement literature.
The SPACE framework~\cite{forsgren2021space} includes efficiency as one of
its five dimensions, defined specifically as the ability to complete work or
make progress on it with minimal interruptions or delays, whether
individually or through a system; the framework treats this dimension as
distinct from \textit{performance} (the outcome achieved) and from
\textit{activity} (the amount of actions performed).
In the Lean and flow management literature, this same angle is
operationalized through the \textit{flow efficiency} metric, defined as the
ratio between the time a task receives active work and the total elapsed
time until its completion~\cite{reinertsen2009principles}; tasks with low
\textit{flow efficiency} spend most of their time waiting, for reviews,
approvals, or dependencies, rather than in execution, regardless of how much
work is eventually delivered.

From the developers' perspective, efficient work involves progressing on and completing relevant tasks with limited friction \cite{meyer2014software,noda2023devex}. Clear goals and requirements help developers direct their effort toward the expected outcome, while interruptions and changes that require a new line of reasoning increase the cost of completing the work \cite{meyer2014software}. Efficiency thus depends not only on a task being completed, but also on how easily developers can translate their effort into meaningful outcomes.

This relationship becomes more evident through feedback loops, cognitive load, and flow state \cite{noda2023devex}. Slow builds, lengthy reviews, insufficient documentation, complex systems, and organizational dependencies consume time and mental effort without necessarily adding value to the final outcome \cite{noda2023devex}. In contrast, fast feedback, clear information, adequate tools, and fewer interruptions allow developers to generate value with less friction and effort \cite{noda2023devex}.

Efficiency should therefore not be treated as a synonym for speed, but
rather as the cost angle, in time, effort, and friction, under which the
same work that productivity describes in terms of achieved outcome can be
analyzed \cite{coelho2025software}.
Developers may perceive themselves as efficient when they achieve the
expected progress within an adequate amount of time, while preserving
quality and avoiding unnecessary effort \cite{coelho2025software}.
This conceptual distinction, however, is not always reflected in the
terminology used by empirical studies: Wagner and Ruhe~\cite{wagner2008systematic},
in a systematic review of productivity factors, observe that productivity,
\textit{performance}, and efficiency are frequently used more or less
synonymously in the Software Engineering literature.
As a result, much of the empirical literature reviewed in this and the
following sections measures productivity directly, without isolating a
separate efficiency metric; when this paper extracts implications for
efficiency from those studies, it refers specifically to the cost and
effort-to-outcome angle of those findings, and not to a measure of
efficiency as defined by the original authors themselves.
This distinction matters for the present research because it directs the
analysis toward the conditions that make software work more difficult,
costly, or slow, even when developers eventually complete their tasks.

\subsection{Longitudinal Monitoring of Developer Efficiency and Productivity}

Developer efficiency and productivity are dynamic phenomena, influenced by technical, organizational, cognitive, and social factors. Developers' perceptions of productivity, as reported in the literature, are shaped by interruptions, communication, tools, code quality, priorities, workflow, and the ability to make meaningful progress \cite{meyer2014software,meyer2019,noda2023devex,cheng2022improves}; under the cost and effort angle established in Section~\ref{sec:background}, these same factors also determine perceived efficiency. Assessments conducted at a single point in time therefore offer only a partial view of both developer productivity and efficiency.

Cross-sectional studies have identified several factors that affect software development productivity, including professional experience, skills, motivation, training, team size, stakeholder involvement, rework, technical knowledge, and organizational processes \cite{canedo2019factors,razzaq2024systematic}. However, because these studies generally collect data at a single point in time, they do not allow one to determine whether such factors are temporary, recurring, or persistent \cite{lynn2009methods,d2024measuring}.

Longitudinal monitoring seeks to overcome this limitation by collecting repeated observations over time. This approach makes it possible to analyze how work conditions and perceived barriers change, persist, or disappear at different points in the development process \cite{lynn2009methods,d2024measuring}. In Software Engineering, this matters because developers' work is continuously affected by changes in requirements, priorities, tools, dependencies, and organizational processes \cite{meyer2019,noda2023devex}.

Despite this, many longitudinal studies use fixed instruments defined before data collection begins. Although fixed instruments favor data comparability, they can limit the identification of new factors that emerge during the study, such as changes related to the use of new tools or generative artificial intelligence \cite{dangelo2024measuring,coutinho2024role}. This limitation motivates adaptive approaches that combine recurring measurements with mechanisms for reviewing the instrument over time.

\subsection{Design Science Research and Action Design Research}

Design Science Research (DSR) is a research approach focused on creating and evaluating artifacts intended to solve relevant problems in organizational and technological contexts \cite{hevner2004design,peffers2007design}. These artifacts can take different forms, such as constructs, models, methods, frameworks, instantiations, or design principles \cite{Wieringa2014,gregor2020research}.

Peffers et al. \cite{peffers2007design} propose a DSR process composed of six activities: problem identification, definition of solution objectives, design and development, demonstration, evaluation, and communication. In this study, DSR provides the overall framework for designing and evaluating the Adaptive Developer Efficiency Monitoring Method (ADEMM).

Action Design Research (ADR) complements DSR by emphasizing the development of the artifact within its actual context of use \cite{Sein2011}. Rather than designing an artifact separately and evaluating it only at the end, ADR proposes iterative cycles of building, intervention, and evaluation, in which the artifact is built, applied, evaluated, and refined in collaboration with organizational actors \cite{Sein2011}.

In this study, ADEMM is treated as the main artifact, while the surveys and interview scripts are considered situated instantiations of the method. This distinction matters because the contribution of the work is not limited to a data collection instrument, but involves an adaptive monitoring method and the design principles derived from its iterative development \cite{Wieringa2014,gregor2020research}.

\subsection{Related Work}

As discussed in Section~\ref{sec:background}, much of the empirical
literature reviewed in this subsection measures productivity directly,
rather than efficiency as defined earlier.
This subsection reports the original findings of these studies in terms of
productivity, which is what they actually measured, and reserves for the
final paragraph the joint implication of these findings for the cost and
effort angle that characterizes efficiency.

Previous studies have investigated factors that affect developer productivity. Canedo and Santos \cite{canedo2019factors} identified productivity factors related to people, product, organization, and open source projects. Meyer et al. \cite{meyer2014software, meyer2019} showed that developers associate productivity with progress, focus, task completion, interruptions, and doing meaningful work.

Other studies analyzed technical and organizational conditions that influence productivity. Cheng et al. \cite{cheng2022improves} showed that code quality and organizational factors affect perceived productivity, while Razzaq et al. \cite{razzaq2024systematic} synthesized factors related to developer experience, including available resources, knowledge, interruptions, code complexity, task context, and standardization. At the team level, Dutra et al. \cite{dutra2015high} conducted a systematic literature review on high-performing teams in Software Engineering, identifying organizational and process factors that favor or hinder such performance; these findings complement the individual factors discussed in this section, but, like the others, were derived from cross-sectional instruments and do not address how such factors evolve over time within the same monitored group.

Longitudinal research has also contributed to this area. D'Angelo et al. \cite{d2024measuring} used recurring surveys to monitor developer experience over time, demonstrating the value of repeated measurements. However, approaches of this kind generally rely on previously defined instruments, which can limit the incorporation of factors emerging during the study \cite{cheng2022improves, lynn2009methods}

Recent research on generative artificial intelligence indicates that developer productivity can also be affected by new and emerging factors. Coutinho et al. \cite{coutinho2024role} found that generative AI can support development activities, but also introduces challenges related to reliability, prompt formulation, validation, and evaluation of generated outputs.

Broad frameworks such as SPACE~\cite{forsgren2021space}, discussed in
Section~\ref{sec:background}, contribute by proposing dimensions that
capture the multifaceted nature of developer productivity, including the
efficiency and flow dimension itself.
ADEMM differs from SPACE in nature, not in content: SPACE is a descriptive
framework, a taxonomy of \textit{what} to observe at one or multiple
measurement instants; ADEMM is an operational method, which specifies
\textit{how} a collection instrument should evolve across successive
iterations, which criteria determine the removal or inclusion of items, and
how to conduct this process when the organization responsible for
monitoring has no employment relationship with the professionals being
observed.
SPACE does not prescribe any of these three elements; ADEMM was built
precisely to fill them in.
For this reason, SPACE and ADEMM operate at complementary, not competing,
levels: the former can inform which dimensions to instantiate within
ADEMM's quantitative channel (Section~\ref{sec:ademm-artifact}), while the
latter governs how that channel evolves over time.

The productivity literature reviewed in this subsection, although not
directly concerned with efficiency, was the empirical foundation that guided
the construction of the instrument items in ADEMM's first
iterations (Section~\ref{sec:ademm-artifact}): factors such as
interruptions, rework, cross-team dependencies, and technical knowledge,
identified by these studies as determinants of productivity, inspired items
that specifically capture their cost and friction dimension, rather than the
aggregate outcome of the work.
This is precisely where this study diverges from the reviewed literature:
while previous works ask how much was delivered, ADEMM asks how much
effort, time, and friction were required to deliver it, and does so
continuously and adaptively.
The gap that motivates ADEMM is therefore not which factors to observe; the
productivity literature already identifies these consistently and served as
the \textit{kernel theory} for the first version of the instrument
(Section~\ref{sec:problem-formulation}).
The gap lies in how to track these same factors, under the efficiency angle,
continuously and adaptively, when the organization responsible for
monitoring does not employ the professionals being observed.

Compared with previous work, this study contributes at the methodological level by proposing ADEMM, an adaptive longitudinal monitoring method. ADEMM differs from previous approaches by treating monitoring as an iterative design process, combining quantitative surveys, qualitative interviews, and evaluation with the problem owner to refine the instrument over time \cite{Sein2011,gregor2020research}.

\section{Research Method}
\label{sec:method}

This study adopted a \textit{Design Science Research} (DSR) approach
combined with \textit{Action Design Research} (ADR)~\cite{Sein2011} to
design, apply, and refine ADEMM over twelve collection cycles.
This combination was motivated by the characteristics of the problem: the
organization needed an artifact that would evolve in response to the
evidence produced during the investigation itself, which required both the
evaluation and communication structure of DSR~\cite{peffers2007design} and
the iterative, situated character of ADR's \textit{Building-Intervention-Evaluation}
(BIE) cycles~\cite{Sein2011}.

\begin{figure*}[t]
    \centering
    \includegraphics[width=\textwidth]{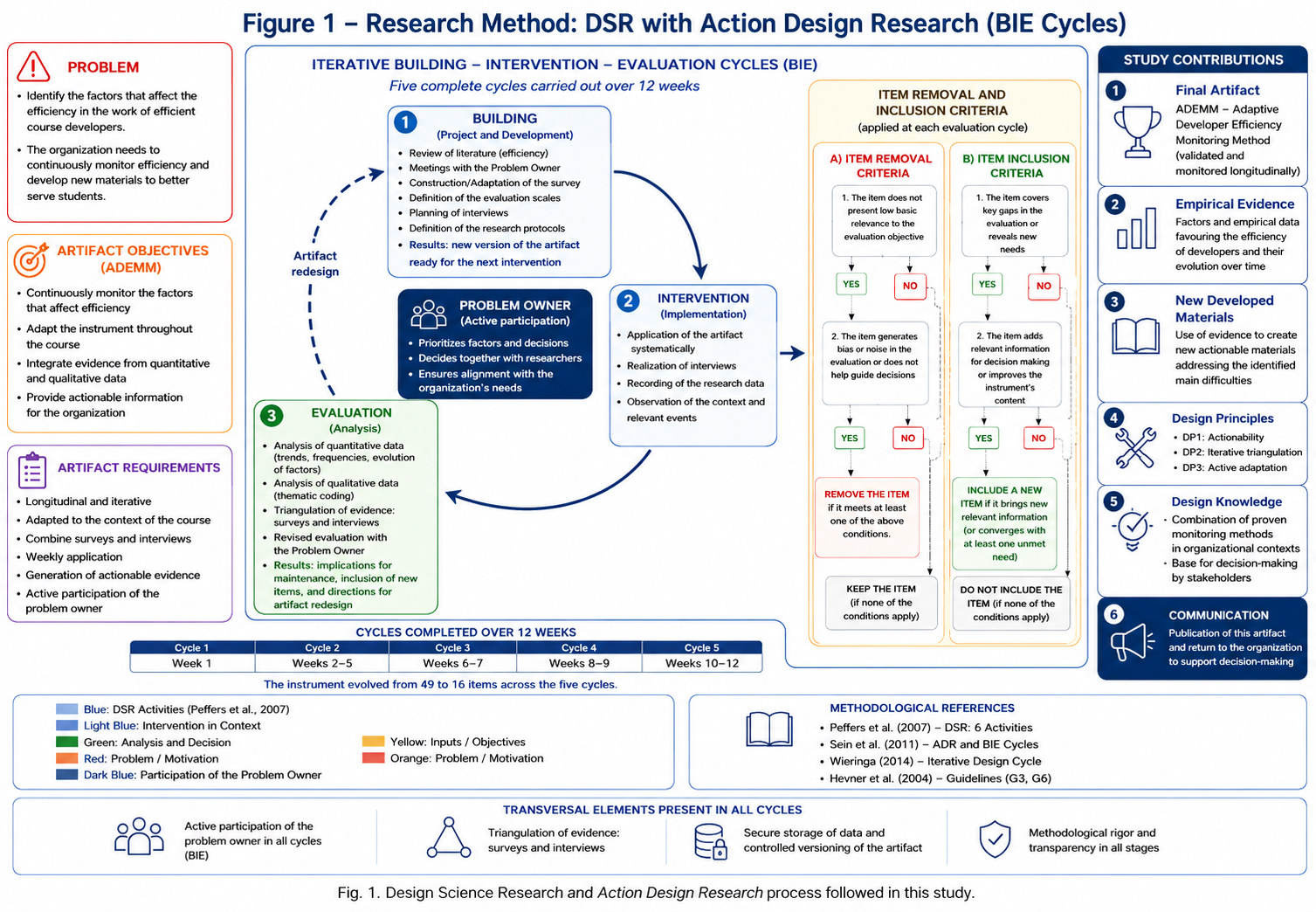}
    \caption{Design Science Research and Action Design Research process followed in the study.}
    \label{fig:dsr-adr-process}
\end{figure*}

The research process is summarized in
Figure~\ref{fig:dsr-adr-process}, which presents the six DSR
activities~\cite{peffers2007design} articulated with ADR's BIE cycles.
The design and development, demonstration, and evaluation activities were
repeated across five design iterations, each corresponding to a complete BIE
cycle conducted in collaboration with the person responsible for the
\textit{Dev Eficiente} consulting practice.
Wieringa~\cite{Wieringa2014} grounds this cyclical nature by distinguishing
problem investigation, treatment design, and treatment validation as
activities that can be repeated before empirical field evaluation.
The communication activity, corresponding to the sixth DSR step, is
materialized through the publication of this paper.

\subsection{Problem Formulation}
\label{sec:problem-formulation}

The project was motivated by the need of an education consulting company,
which follows developers affiliated with different client companies, to
understand the factors affecting these professionals' efficiency during
their work.
Unlike the more common context in the literature, in which the organization
conducting the research directly employs the observed
developers~\cite{meyer2019,cheng2022improves}, this company had no
employment relationship with, nor operational control over, the
participants.
Its interest was instrumental: understanding which barriers affected the
efficiency of its students in order to guide the content, mentoring, and
technical training offered by the \textit{Dev Eficiente} consulting
practice.
This structural constraint is not exclusive to the investigated consulting
firm: coding bootcamps and technology schools, corporate upskilling
programs, consulting firms that place developers in client squads, and
on-demand staffing platforms all face the same limitation of needing to
monitor professionals' efficiency without holding operational control over
the environment in which those professionals actually work.

Throughout this paper, \textit{problem owner} designates the organizational
lead responsible for the \textit{Dev Eficiente} consulting practice, who
participated in all evaluation meetings at the end of each BIE cycle and had
the authority to decide on the retention, removal, or inclusion of
instrument items, as well as to direct the practical use of the findings in
the consulting practice's content, mentoring, and technical training
decisions.
This role corresponds to the organizational actor role anticipated by
ADR~\cite{Sein2011} in \textit{organization-dominant} BIE cycles: someone
with visibility into the organization's current priorities and the ability
to translate research findings into action, but who does not take part in
the technical analysis of the data.

The organizational problem preceded any choice of instrument or technology,
which corresponds to the \textit{organization-dominant} BIE cycle profile
described by Sein et al.~\cite{Sein2011}.
The collection instrument was therefore a consequence of the problem
formulation, not its starting point.
The literature review on factors affecting developer
efficiency~\cite{cheng2022improves,razzaq2024systematic} served as the
empirical basis for the first version of the instrument, functioning as
\textit{kernel theory} in the sense accepted by ADR: prior empirical
generalizations about the phenomenon, not necessarily formal explanatory
theories.

The problem identified was the absence of a structured, continuous process
capable of identifying, over time, which factors affected the efficiency of
developers outside the organization's direct control, and of translating
this evidence into practical guidance for the consulting practice.
This problem was broken down into four operational objectives,
corresponding to the second DSR activity~\cite{peffers2007design}:

\begin{enumerate}[label=(\roman*)]
    \item continuously, rather than at a single point in time, identify the
    factors undermining participants' efficiency;

    \item keep the instrument adaptable to changes in the consulting
    practice's content and activities over time;

    \item triangulate quantitative and qualitative evidence, so as to
    capture both the frequency and the context of each factor;

    \item produce actionable information, that is, linked to concrete
    decisions about content, mentoring, or technical training tracks, for
    the person responsible for the consulting practice, without depending
    on direct access to the companies where the participants worked.
\end{enumerate}

\subsection{Context and Participants}
\label{sec:participants}

The research was conducted between February and May 2026 within the
\textit{Dev Eficiente} consulting practice.
Eligible participants were developers already served by the consulting
practice, with no external recruitment or additional referral process.
In total, 27 developers were followed throughout the study; of these, 18
took part in the semi-structured interviews.
Table~\ref{tab:participant-seniority} presents the distribution by
seniority level.

\begin{table}[ht]
\centering
\caption{Distribution of participants by seniority level.}
\label{tab:participant-seniority}
\begin{tabular}{lcc}
\toprule
\textbf{Seniority level} & \textbf{n} & \textbf{\%} \\
\midrule
Senior & 12 & 44.4 \\
Mid-level  & 10 & 37.0 \\
Junior &  5 & 18.5 \\
\midrule
\textbf{Total} & \textbf{27} & \textbf{100.0} \\
\bottomrule
\end{tabular}
\end{table}

Participants worked at distinct companies and teams, with no organizational
tie in common beyond their relationship with the consulting practice, which
characterizes the panel as heterogeneous with respect to respondents' work
contexts.
This heterogeneity is consistent with the method's objective: ADEMM needs to
be applicable regardless of each participant's specific environment, given
that the organization has no access to those environments.

All participants were informed of the research objectives and formally
consented to the collection and use of the data through an Informed Consent
Form.

\subsection{Data Collection}
\label{sec:data-collection}

The study adopted a panel design~\cite{lynn2009methods}, with the same
sample of 27 participants followed across twelve collection cycles.
Surveys were administered weekly between weeks 1 and 7 and biweekly between
weeks 8 and 12, totaling approximately four months of monitoring.
The sample remained stable in terms of participant identity throughout the
study; there was, however, intermittent non-response across cycles.
The number of weekly respondents ranged between 13 and 25 of the 27
participants, a pattern consistent with the literature on longitudinal
panels, which points to occasional non-response as a common phenomenon in
the absence of definitive attrition~\cite{lynn2009methods}.

Each collection cycle included a quantitative-qualitative survey composed of
binary items (Yes/No), scale items, and two open-ended questions.
The binary items covered friction factors identified in the literature and
emerging from the data; the open-ended questions asked participants to
describe the main factor perceived that week and to indicate anything
relevant not covered by the closed items.
In total, the instrument went through five versions throughout the study,
as detailed in Section~\ref{sec:ademm-artifact}.

The semi-structured interviews were conducted in three rounds throughout the
collection period, with a subset of 18 participants.

\subsection{Data Analysis}
\label{sec:data-analysis}

The quantitative survey data were analyzed using descriptive statistics,
including response frequencies per item and per collection cycle, and
mixed-effects models to separate between-participant variation from
within-participant variation over time.

The qualitative data from the open-ended questions were analyzed using
inductive thematic analysis~\cite{braun2006using}.
The themes identified in each collection cycle were discussed in joint
review meetings between the researchers and the person responsible for the
consulting practice (\textit{problem owner}), who assessed the practical
relevance of the emerging patterns for decisions about the instrument and
for planning the consulting practice's activities.

Triangulation between quantitative and qualitative data occurred at two
points: during each BIE cycle, in the assessment of response patterns and
participant narratives that guided instrument revisions; and at the end of
the study, in the interpretation of the design principles derived from the
iterative process, as described in Section~\ref{sec:results}.
\section{The Artifact: ADEMM}
\label{sec:ademm-artifact}

The developed artifact, called the \textit{Adaptive Developer Efficiency
Monitoring Method} (ADEMM), does not correspond to a single, isolated
collection instrument.
It is an adaptive longitudinal monitoring method aimed at continuously
identifying the factors that affect developer efficiency in contexts where
the organization responsible for the follow-up has no employment
relationship with, or operational control over, the professionals being
monitored.

ADEMM is composed of three integrated components.
The first is a recurring quantitative channel, composed of surveys with
closed binary items and open items, applied weekly or biweekly to
participants.
The second is a complementary qualitative channel, composed of
semi-structured interviews conducted in cycles, whose findings directly feed
the revision of the quantitative channel.
The third is a joint review process with the \textit{problem owner}, which
takes place at the end of each iteration and applies explicit criteria to
decide which items to keep, remove, or include in the next iteration.
The integration of these three components is what distinguishes ADEMM from
a conventional fixed-instrument longitudinal survey.

The collection instruments, including the survey versions and interview
scripts, constitute situated instantiations of ADEMM at each design
iteration, rather than the artifact itself.
Gregor et al.~\cite{gregor2020research} distinguish \textit{situated
implementations}, useful to the specific context in which they were
applied, from \textit{emerging design theories}, which express principles
transferable to other contexts.

\subsection{Instrument Evolution Criteria}
\label{sec:evolution-criteria}

The evolution of the instrument between iterations followed explicit
criteria, applied by consensus between the researchers and the person
responsible for the consulting practice in evaluation meetings at the end of
each BIE cycle.

Removal of an item was considered when two signals converged.
The first was low variance in the closed-item responses across the weeks of
the iteration, that is, when an item received predominantly the same
response, indicating that the corresponding barrier was not occurring
variably among participants.
The second was the \textit{problem owner}'s assessment of the item's
usefulness for guiding decisions that support participants' efficiency, such
as content, mentoring, and technical training, within the context of the
\textit{Dev Eficiente} consulting practice.
When both signals converged, the item was removed in the subsequent
iteration.
Convergence of the two criteria was necessary because low-variance items can
simply be low-prevalence questions that remain informative, while items
judged as not very useful by the \textit{problem owner} may show high
variance without generating organizational action.

One example: the item "This week, the work process (Scrum, Kanban, etc.)
hindered the delivery flow" showed between 3.7\% and 11.1\% affirmative
responses in Weeks~2 through~5, and was assessed by the person responsible
for the consulting practice as having limited usefulness for guiding
decisions that support participants' efficiency.
With both criteria satisfied, the item was removed starting from
Iteration~3.

The inclusion of new items followed the reverse process: recurring patterns
identified in the interview analysis were brought to the evaluation meetings
and, when confirmed as relevant to the consulting practice's context, gave
rise to new closed items in the subsequent iteration.
The clearest example was the inclusion of items about artificial
intelligence in Survey~V: the theme had spontaneously emerged in the
interviews as a source of difficulties and as a potential strategy for
reducing dependence on other areas, one of the most persistent factors
observed in the surveys.

Before and during the first iterations, the researchers considered
alternative forms of collection beyond surveys and interviews.
A work diary was introduced as a complementary instrument in
Iterations~2 and~3.
The instrument allowed participants to freely record, throughout the week,
situations related to efficiency in their activities.
However, the diary received only a single response during the entire period
in which it was available and was discontinued starting from Iteration~4.
Low adherence is a known risk in diary methods, due to the ongoing effort of
self-reporting~\cite{bolger2003diary}, compounded by the intense work
routine of the participating developers.
The attempt with the diary constitutes evidence of an effective cycle of
generating and testing a design alternative: its discontinuation was a
data-driven decision, not an a priori judgment.
Other alternatives, such as tool usage telemetry and direct observation,
were discarded already at the design stage because they would require
access to the infrastructure of the participants' employing companies,
organizations outside the consulting practice's scope of control.

\subsection{Instantiating ADEMM: Five Iterations at Dev Eficiente}
\label{sec:bic-iterations}

ADEMM does not prescribe a fixed number of iterations, a specific set of
factors to monitor, or a standardized instrument.
What the method defines are the evolution criteria described in
Section~\ref{sec:evolution-criteria} and the joint review process with the
\textit{problem owner}.
The concrete elements of each instantiation, including which factors to
investigate, over how many weeks, with which instrument, and in which
organizational context, are local decisions that each organization makes
when instantiating the method.

This subsection documents the first instantiation of ADEMM, conducted within
the \textit{Dev Eficiente} consulting practice over twelve collection
cycles.
The specific decisions described here, such as the weekly cadence, the
binary items, and the survey themes, are not part of the method itself:
they are part of the example.
What varies between instantiations are the local decisions: the number of
cycles, the cadence, the item format, the factors investigated, and the
participant profile.
What remains constant is the logic of the process: a recurring quantitative
channel, an integrated qualitative channel, joint review with the
\textit{problem owner}, and explicit criteria for evolving the instrument.

Table~\ref{tab:design-iterations} summarizes the five iterations, including
the number of closed items in each version of the instrument.

\begin{table*}[t]
\caption{Design iterations of the monitoring instrument}
\label{tab:design-iterations}
\centering
\footnotesize
\renewcommand{\arraystretch}{1.15}
\begin{tabularx}{\textwidth}{p{1.8cm} p{2.2cm} p{1.2cm} X}
\toprule
\textbf{Iteration} & \textbf{Weeks} & \textbf{Closed items} & \textbf{Main changes relative to the previous version} \\
\midrule

Survey I  & Week 1   & 49 &
Initial version, with Likert-scale items organized into twelve factor groups,
including planning, technical problem solving, decision making,
communication, and dependencies with other areas. \\

Survey II & Weeks 2--5 & 19 &
Reduction and reorientation of the investigated factors, focusing on
prioritization and clarity of work, technical knowledge, autonomy, external
dependencies, work process, and collaboration. Replacement of the Likert
scale with binary (Yes/No) items and inclusion of two open-ended questions. \\

Survey III & Weeks 6--7 & 17 &
Removal of two items with low variance and low usefulness for guiding
efficiency support actions (Scrum/Kanban work process; psychological
safety). Simplification of the wording, eliminating the repeated expression
"this week" in the statements. \\

Survey IV & Weeks 8--9 & 17 &
No content changes. The decision not to alter the instrument was itself a
result of the evaluation, confirming the adequacy of the investigated
factors for the period. \\

Survey V  & Weeks 10--12 & 16 &
Removal of four items with decreasing variance and reduced relevance for the
final period of the follow-up. Inclusion of three items on the use of
artificial intelligence in development activities, motivated by an emerging
qualitative signal from the interviews. \\

\bottomrule
\end{tabularx}
\end{table*}

\paragraph{Iteration 1 - Survey I.}
Based on the kernel theory identified in the
literature~\cite{cheng2022improves,razzaq2024systematic} and a prioritization
meeting with the person responsible for the consulting practice, Survey~I
was developed with 49 Likert-scale questions, organized into twelve factor
groups.
The evaluation after Week~1 identified the most practically relevant groups
and showed that the length of the instrument and the Likert format were
unsuitable for weekly collection with active developers.

\paragraph{Iteration 2 - Survey II.}
Based on the evaluation of Iteration~1, the instrument was restructured:
the Likert scale was replaced with binary (Yes/No) items, the number of
factors was reduced, and two open-ended questions were incorporated to
capture context not covered by the closed items.
The instrument was applied weekly over four weeks.
The periodic evaluations indicated the need to simplify the wording,
eliminating the repetition of the expression "this week" in the statements.
The qualitative channel, composed of eight interviews conducted during this
period, identified two emerging themes that would require further
exploration: the use of artificial intelligence and process-related
barriers to workflow.

\paragraph{Iteration 3 - Survey III.}
The instrument was simplified to 17 closed items and two open-ended
questions, with more direct wording.
Two items were removed for simultaneously satisfying the criteria of low
variance and low usefulness for guiding efficiency support actions, as
detailed in Section~\ref{sec:evolution-criteria}.
The evaluation after Weeks~6 and~7 confirmed the adequacy of the revised
structure.

\paragraph{Iteration 4 - Survey IV.}
No items were changed.
The stability of the instrument preserved the comparability of results
across the weeks in this period and confirmed that the investigated factors
remained adequate to the participants' activities.

\paragraph{Iteration 5 - Survey V.}
The analysis of the results from Iteration~4, combined with a review of the
consulting practice's content portfolio and the emerging themes from the
second round of interviews, motivated two changes.
Four items were removed due to decreasing variance and reduced relevance for
the final period: waiting for external validations, definition of task
completion criteria, adequacy of technical complexity to available time, and
lack of direction from leadership.
Three items on the use of artificial intelligence were incorporated,
directly responding to the qualitative signal identified in the Iteration~2
interviews.
The new items captured an emerging factor not anticipated in the initial
problem formulation, demonstrating the method's ability to incorporate
unanticipated dimensions without a complete redesign of the instrument.

\subsection{Qualitative Cycles}
\label{sec:qualitative-cycles}

The qualitative channel operated across two cycles of semi-structured
interviews, integrated with the instrument review process.

In the first cycle, composed of eight interviews conducted during
Iteration~2 (Weeks~2 to~5), the script investigated general factors related
to efficiency, including activity organization, requirements, estimates,
and tools used.
The analysis of this cycle identified two themes that were not adequately
covered by the closed items in force at the time: the use of artificial
intelligence and process-related barriers to workflow.
These findings were presented to the \textit{problem owner} in the
Iteration~2 evaluation and influenced both the simplification of Survey~III
and the definition of the scope for the second round of interviews.

In the second cycle, composed of ten interviews conducted during
Iterations~3 and~4, the script explored in greater depth the two themes
identified in the first cycle.
The findings from this cycle confirmed the relevance of artificial
intelligence use as both a support and friction factor in development
activities, providing the basis for the inclusion of the three
corresponding items in Survey~V.

The interview data were analyzed using deductive and inductive thematic
analysis~\cite{braun2006thematic}.
The deductive analysis was guided by the factors investigated in the surveys
and by the topics of the scripts; the inductive analysis allowed the
identification of unanticipated emerging patterns.
After full reading of the transcripts, coding, and iterative grouping of the
codes, the 236 identified codes were organized into seven analytical themes,
presented in Table~\ref{tab:rq3-themes}.

\begin{table*}[t]
\caption{Analytical themes identified in the semi-structured interviews}
\label{tab:rq3-themes}
\centering
\footnotesize
\renewcommand{\arraystretch}{1.15}
\begin{tabularx}{\textwidth}{
>{\RaggedRight\arraybackslash}p{5.2cm} Y}
\toprule
\textbf{Theme} & \textbf{Description} \\
\midrule

Dependencies, communication, and collaboration &
Need for interaction, alignment, or support from other people, teams,
areas, or stakeholders to carry out activities. \\

Organizational processes, planning, and prioritization &
Organization of activities, internal workflows, priority definition, effort
estimation, and practices that guide the work. \\

Understanding of the business, system, and requirements &
Understanding of business rules, existing systems, requirements, and the
information needed to develop a solution. \\

Interruptions and focus at work &
Situations that interrupt reasoning, hinder concentration, or increase
mental load during development. \\

Technical knowledge &
Mastery of technologies, tools, components, practices, and architectural
decisions required to perform the tasks. \\

Use of AI as support and as a source of challenges &
Situations involving the use of artificial intelligence tools in
understanding, implementation, debugging, code generation, and
decision-making support activities. \\

Quality, validation, and rework &
Practices and conditions related to testing, code review, delivery
validation, defect identification, and problem correction. \\

\bottomrule
\end{tabularx}
\end{table*}

\subsection{Artifact Evaluation}
\label{sec:artifact-evaluation}

The evaluation of ADEMM emerges from the consolidation of the formative
evaluations conducted at the end of each BIE cycle, following the principle
of \textit{guided emergence} proposed by Sein et al.~\cite{Sein2011}: design
knowledge emerges from the continuous interaction between intervention in
the artifact and the organization's reaction over the cycles, rather than
from a single evaluation event at the end of the study.

The formative evaluations and the consolidated summative evaluation applied
four criteria derived from the operational objectives defined in
Section~\ref{sec:problem-formulation}.
Table~\ref{tab:artifact-evaluation-criteria} presents these criteria and
the corresponding evidence.

\begin{table*}[t]
\caption{Artifact evaluation criteria and observed evidence}
\label{tab:artifact-evaluation-criteria}
\centering
\footnotesize
\renewcommand{\arraystretch}{1.15}
\begin{tabularx}{\textwidth}{p{3.3cm} p{4.0cm} X}
\toprule
\textbf{Criterion} & \textbf{What it evaluates} & \textbf{Observed evidence} \\
\midrule

Ability to identify relevant and emerging factors &
Verifies whether the instrument captured both anticipated and unanticipated
factors. &
Identification of 236 distinct codes and seven analytical themes in the
interviews; incorporation of the artificial intelligence theme in
Iteration~5 based on an emerging qualitative signal. \\

Adaptability to contextual changes &
Verifies whether the instrument could be adjusted without compromising the
continuity of collection. &
Five iterations with documented decisions to retain, remove, and include
items; discontinuation of the work diary after low adherence without
interrupting the main collection. \\

Practical usefulness for the organization &
Verifies whether the results generated concrete actions, rather than merely
a descriptive report. &
Creation of a new course focused on requirements analysis and improvement,
still during the execution of the study, in response to two persistent
findings: dependence on other areas and a gap in business and requirements
understanding. \\

Comparability of results across iterations &
Verifies whether the changes preserved the ability to compare trends over
time. &
Full retention of the Survey~III structure in Iteration~4 with no content
changes; only textually stable items across iterations support direct trend
comparisons. \\

\bottomrule
\end{tabularx}
\end{table*}

The creation of the new course from intermediate findings, before the
study's conclusion, illustrates the co-evolution between artifact and
organization characteristic of \textit{organization-dominant} BIE cycles:
the knowledge generated by the iterations influenced organizational
decisions during the process, not only after its conclusion.

\subsection{Formalizing the Design Principles}
\label{sec:design-principles}

At the end of the five iterations, the researchers and the \textit{problem
owner} jointly reviewed the history of decisions to retain, remove, and
include items in light of the four evaluation criteria.
The goal was to extract generalizable lessons about the adaptive
longitudinal monitoring process, not only about the efficiency factors
observed in this instantiation.

This process resulted in the formalization of three design principles,
expressed according to the context, intervention, mechanism, and expected
outcome structure~\cite{gregor2020research}.

\vspace{0.4em}
\noindent\textbf{DP1. Joint prioritization with the \textit{problem owner},
guided by actionability.}

In contexts where the organization responsible for the follow-up has no
direct employment relationship with, or operational control over, the
professionals being monitored (\textit{context}), the investigated factors
should be jointly reviewed and prioritized with the organizational lead at
the end of each iteration, favoring factors over which the organization has
real capacity to intervene (\textit{intervention}).
This principle is necessary because the \textit{problem owner} has
visibility into current activities and priorities to which the researchers
have no direct access, and because instruments that capture non-actionable
barriers produce diagnoses without effective capacity for
response (\textit{mechanism}).
The expected outcome is an instrument permanently aligned with the
organization's operational needs and capacity for action, rather than only
with the kernel theory that guided its first version (\textit{outcome}).

\vspace{0.4em}
\noindent\textbf{DP2. Combination of closed items and a qualitative channel
per iteration.}

When the monitored phenomenon is dynamic and cannot be fully anticipated at
the start of the study (\textit{context}), each iteration of the instrument
should combine closed items, which allow tracking the frequency and
persistence of already known factors, with open items and interviews, which
make it possible to capture context and emerging factors
(\textit{intervention}).
Closed items alone do not capture emerging factors; open items alone do not
allow quantifying the persistence of factors over time; the qualitative
channel feeds the redesign of the quantitative
channel (\textit{mechanism}).
The expected outcome is the identification of emerging factors, such as the
use of artificial intelligence in this study, without requiring a complete
redesign of the instrument at each iteration (\textit{outcome}).

\vspace{0.4em}
\noindent\textbf{DP3. Removal by low variance and inclusion by emerging
signal.}

When the instrument accumulates items across several
iterations (\textit{context}), items with low response variance and
decreasing usefulness as recognized by the \textit{problem owner} should be
removed by consensus; recurring patterns identified in the qualitative
analysis should support the creation of new closed items in subsequent
iterations (\textit{intervention}).
Instrument length has a direct cost on response rate and participant
fatigue~\cite{galesic2009effects}; keeping low-informative-value items
compromises the quality of responses to the items that remain
relevant (\textit{mechanism}).
The expected outcome is an instrument that preserves the comparability of
the stable core items across iterations without growing indefinitely in
length (\textit{outcome}).

These three principles constitute the formalization of the lessons learned
across the five BIE iterations and represent the main design contribution of
this study.
This contribution is distinct from the empirical findings about the factors
affecting the participating developers' efficiency: while the empirical
findings describe what was observed in this instantiation, the design
principles guide how to conduct similar monitoring efforts in other
contexts.
\section{Results}
\label{sec:results}

This section answers the research question by presenting ADEMM in its
abstract form, generalizable to contexts beyond the study that originated
it.
The answer has two parts: the description of the method as an eight-step
process, independent of the investigated domain, and the demonstration of
its transferability through an alternative scenario.
The three design principles underlying these steps were formalized in
Section~\ref{sec:design-principles} through systematic reflection on the
five iterations reported in Section~\ref{sec:ademm-artifact}.

\subsection{ADEMM as a Transferable Process}
\label{sec:ademm-abstract}

\begin{figure}[t]
\centering
\includegraphics[width=\linewidth]{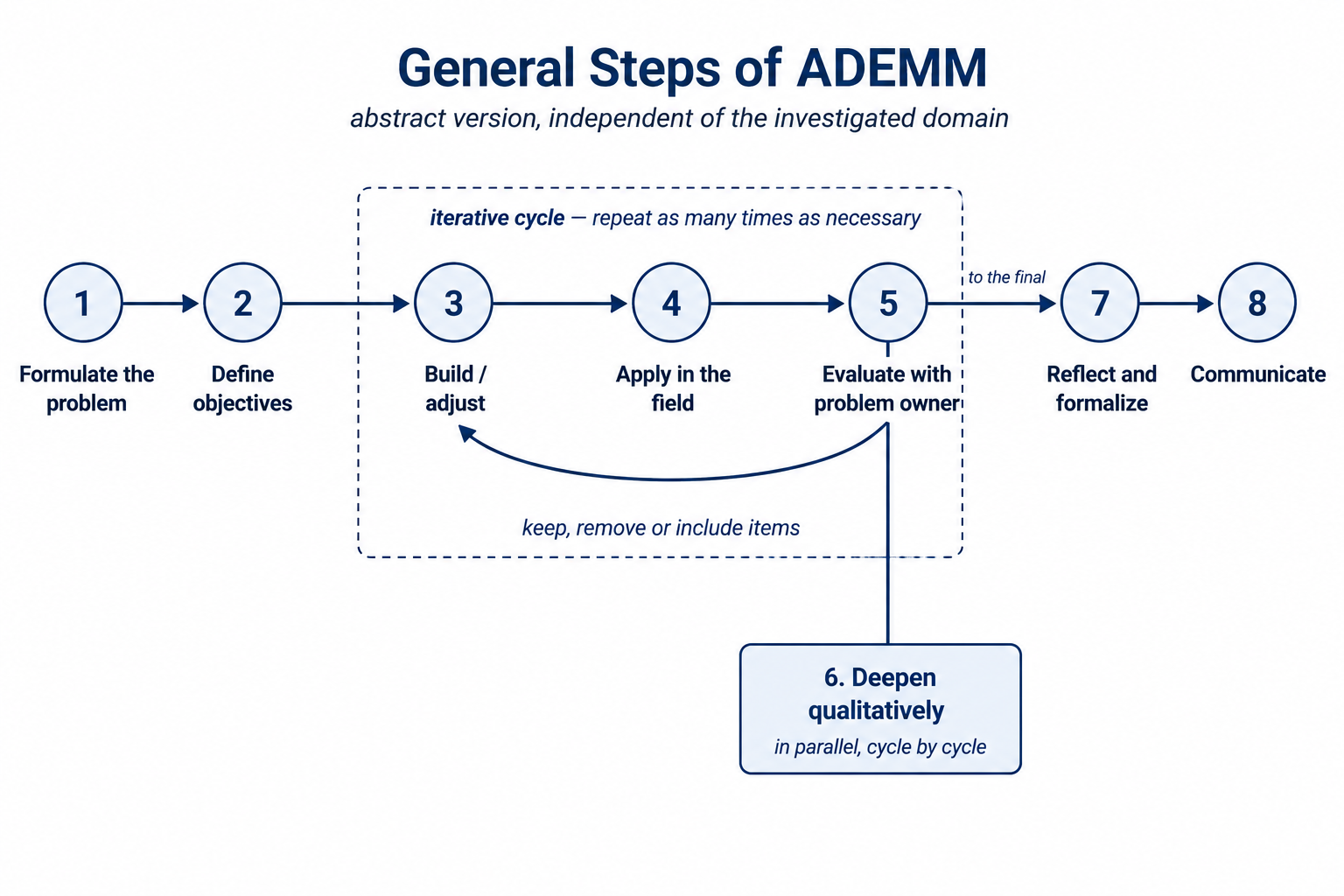}
\caption{Overall flow of the \textit{Adaptive Developer Efficiency
Monitoring Method} (ADEMM) in eight steps, independent of the investigated
domain.}
\label{fig:ademm_overview}
\end{figure}

\begin{figure*}[t]
\centering
\includegraphics[width=\textwidth]{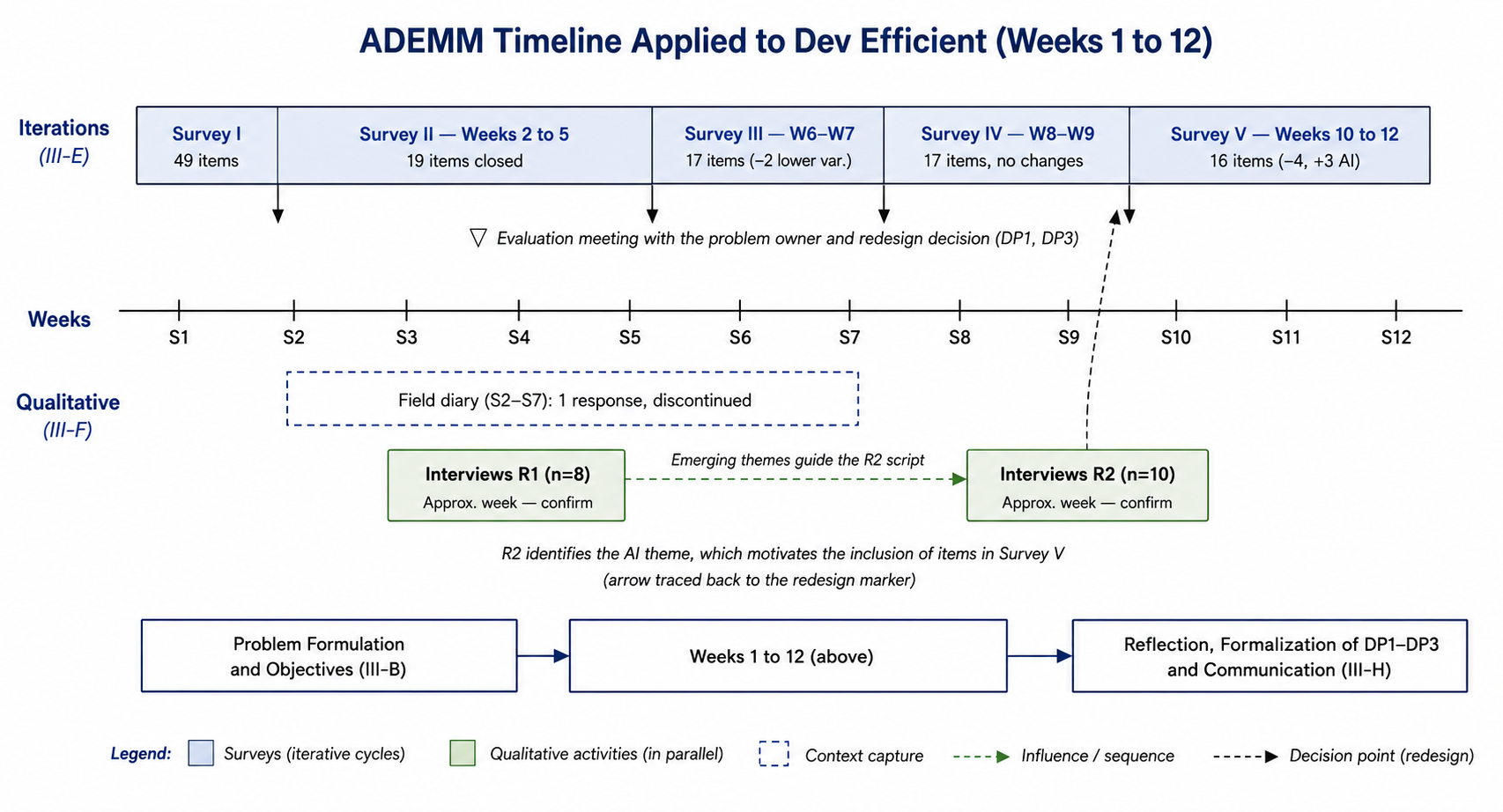}
\caption{Instantiation of ADEMM across the twelve weeks of the
\textit{Dev Eficiente} study. The quantitative survey iterations were
continuously informed by qualitative evidence and redesign meetings with the
\textit{problem owner}.}
\label{fig:ademm_timeline}
\end{figure*}

Figure~\ref{fig:ademm_overview} presents ADEMM as an eight-step process
independent of the investigated domain.
Figure~\ref{fig:ademm_timeline} shows the concrete instantiation of this
process throughout the twelve weeks of the \textit{Dev Eficiente} study.
The distinction between the two figures is deliberate: the first is the
method; the second is one of its possible instantiations.

The eight steps do not constitute an alternative research process proposal
to DSR: they are an operational translation of the six activities of
Peffers et al.~\cite{peffers2007design} that makes the process followable by
the \textit{problem owner} without requiring familiarity with academic
terminology.
Steps~1 and~2 correspond to problem identification and objective
definition; Steps~3 through~5 operationalize design, demonstration, and
evaluation as a repeated cycle, following Sein et al.'s~\cite{Sein2011} BIE
cycles; Step~6 corresponds to the qualitative deepening that feeds the
redesign of the instrument (DP2); Step~7 corresponds to the reflection and
formalization of design knowledge; and Step~8 corresponds to communication.

Table~\ref{tab:passos_ademm} details each step with its contextual
variations, indicating where local decisions replace the choices made in
this study without compromising the logic of the method.

\begin{table*}[htbp]
\caption{General steps of ADEMM and contextual variations}
\label{tab:passos_ademm}
\centering
\footnotesize
\renewcommand{\arraystretch}{1.2}
\begin{tabularx}{\textwidth}{
>{\hsize=0.22\hsize\RaggedRight\arraybackslash}X
>{\hsize=0.38\hsize\RaggedRight\arraybackslash}X
>{\hsize=0.40\hsize\RaggedRight\arraybackslash}X}
\toprule
\textbf{Step} & \textbf{What to do} & \textbf{Contextual variations} \\
\midrule

1. Formulate the problem &
Identify the motivating organizational problem together with the
\textit{problem owner} and gather an initial empirical basis. &
The empirical basis can come from a systematic review, published cases, or
the organization's internal records. \\

2. Define objectives &
Derive explicit operational objectives from the problem, focusing on what
the organization can monitor and influence. &
The scope can be broad (general diagnosis) or narrow (actionable factors),
depending on the organization's capacity to intervene. \\

3. Build the instrument &
Develop closed and open items from the empirical basis, prioritized with
the \textit{problem owner} (DP1). &
The format can vary (Likert, binary, frequency); the number of dimensions
depends on the maturity of the investigated domain. \\

4. Apply in the field &
Apply the instrument directly to participants at a regular cadence. &
The cadence can be weekly, biweekly, or monthly, depending on the phenomenon
and the risk of respondent fatigue. \\

5. Evaluate with the \textit{problem owner} &
Review the cycle's results and decide on retaining, removing, or including
items (DP1, DP3). &
Removal criteria include low variance, redundancy, or direct feedback from
participants on item relevance. \\

6. Deepen qualitatively &
Conduct interviews or another technique in parallel, feeding back into the
instrument (DP2). &
Interviews can be replaced by focus groups or open questions within the
survey itself; diaries require caution regarding adherence. \\

7. Reflect and formalize &
Consolidate the decisions made across cycles into generalizable design
principles. &
Formalization can produce design principles in CIMO format, a practical
checklist, or an operational playbook. \\

8. Communicate &
Report the artifact and the findings to the organization and the academic
community. &
Can take the form of a scientific paper, an internal report, or both. \\

\bottomrule
\end{tabularx}
\end{table*}

\subsection{Transferability of the Method}
\label{sec:transferability}

To make the transferability claim concrete, consider a second education
organization that offers mentoring to product designers working as
freelancers for distinct clients, with no employment relationship with any
of them.
The structure of the problem is analogous: the organization needs to
monitor the efficiency of professionals working in environments outside its
control.
Applying ADEMM would follow the eight steps in
Figure~\ref{fig:ademm_overview}, but with decisions specific to the design
context.

In problem formulation (Step~1), the organization would review the
literature on designer efficiency and identify, together with its
\textit{problem owner}, which factors would be priorities for the mentoring
program.
In building the instrument (Step~3), DP1 would guide the prioritization of
items on which mentoring can act, such as clarity of the brief received
from the client, over factors out of reach, such as the client's internal
organizational culture.
In subsequent iterations, DP2 would guide the maintenance of the
qualitative channel to capture emerging dimensions specific to design work,
and DP3 would guide the removal of low-variance items and the inclusion of
new factors identified in the interviews.

What transfers between the two contexts is not the content of the
instrument: the efficiency factors for software developers and for
freelance designers are distinct, and the corresponding closed items would
be entirely different.
What transfers is the logic of the process: the integration between the
quantitative and qualitative channels, the joint review with the
\textit{problem owner} at each iteration, the explicit criteria for removal
by low variance and inclusion by emerging signal, and the guidance by
actionability in prioritizing factors.

What does not transfer without adaptation are the specific structural
conditions of this study: the weekly cadence may not be adequate for
more slowly changing phenomena; the binary composition of the items may not
capture relevant gradations in other domains; and the number of iterations
needed depends on how quickly the investigated phenomenon changes.
These are the variables that each instantiation of ADEMM must calibrate
locally, within the method's general logic.

\section{Discussion}
\label{sec:discussion}

ADEMM addresses a specific structural problem: how to longitudinally
monitor the efficiency of professionals when the organization responsible
for the follow-up has no access to their work environment.
Section~\ref{sec:results} demonstrated how the method works in the abstract
and how its eight steps can be instantiated in a domain different from the
one that originated it.
This section discusses what ADEMM adds to the DSR and ADR literature, what
the specific limitations of the method identified during its application
are, and what should be examined in future replications.

\subsection{What ADEMM Adds Beyond a Generic Application of DSR}
\label{sec:ademm-contribution}

ADEMM was built entirely within the boundaries of DSR and ADR.
None of its eight steps contradicts or replaces the six activities of
Peffers et al.~\cite{peffers2007design}, the BIE cycles of
Sein et al.~\cite{Sein2011}, or the design cycle of
Wieringa~\cite{Wieringa2014}.
The relevant question, therefore, is not whether ADEMM diverges from DSR,
but which methodological decisions it makes explicit where DSR and ADR leave
choices open for each case.
Table~\ref{tab:ademm-contribution} organizes this difference along the five
main decision points of the method.

\begin{table*}[t]
\caption{ADEMM's contributions relative to a generic application of DSR and ADR}
\label{tab:ademm-contribution}
\centering
\footnotesize
\renewcommand{\arraystretch}{1.15}
\begin{tabularx}{\textwidth}{p{3.0cm} p{4.1cm} X}
\toprule
\textbf{Decision point} &
\textbf{What generic DSR/ADR offers} &
\textbf{What ADEMM adds, tested in this study} \\
\midrule

Item pruning criterion &
Prescribes evaluation cycles (BIE), but does not define when an item should
be removed from the instrument. &
An operational criterion based on two converging signals: low variance in
closed responses and usefulness for guiding efficiency support actions, as
recognized by the \textit{problem owner} (DP3). Applied and documented
across five real iterations. \\

Organization without control over the root cause &
Recognizes \textit{organization-dominant} cycles, but does not specify how
to prioritize factors when the organization has no influence over the
causes of the monitored problems. &
An explicit actionability criterion (DP1), which prioritizes factors over
which the organization can intervene, avoiding diagnoses without capacity
for response. \\

Cadence and collection fatigue &
Does not define the application cadence or discuss fatigue risks in
longitudinal studies. &
A tested cadence, initially weekly and later biweekly. Empirical evidence
that a free-form logging instrument had virtually no adherence in this
population. This negative result was documented to guide future
replications. \\

Handling non-response &
Does not prescribe how to compute indicators when part of the sample does
not respond to a collection cycle. &
An explicitly documented fixed-denominator convention ($N=27$), with its
implications: weeks with higher non-response produce conservatively lower
percentages, which should be considered when interpreting temporal
variations. \\

Format of design knowledge &
Recommends formalizing design knowledge at the end of the study, without
requiring a specific format. &
Formalization of three principles in CIMO format, with explicit
traceability to the decisions made during the iterations, distinguishing
them from generic recommendations with no empirical grounding. \\

\bottomrule
\end{tabularx}
\end{table*}

The contribution of this study lies not in the six-activity structure
itself, which already existed and is fully preserved, but in the
operational refinement that resolves, in a tested and documented way, a set
of decisions that DSR and ADR leave open.
It is this refinement, present in the three design principles and in the
methodological conventions made explicit throughout the paper, that an
organization gains by adopting ADEMM rather than starting from scratch from
Peffers et al. or Sein et al.

\subsection{Process Limitations and Points to Test in Replications}
\label{sec:process-limitations}

The points below do not correspond to threats to the validity of the
empirical findings: they represent weaknesses of ADEMM itself as a design
artifact, identified during its application, that should be tested or
corrected in future instantiations.

The pruning criterion based on low variance, formalized in DP3, favors
frequency over severity.
A factor that affects few participants but has severe consequences when it
occurs could be removed by the same criterion used to discard genuinely
irrelevant items.
A future version of ADEMM should complement variance analysis with a
perceived severity indicator, possibly extracted from open-ended responses
or from a specific scale, before permanently removing an item.

The same low-variance pruning criterion also does not distinguish
irrelevant factors from seasonal or cyclical factors, whose relevance can
temporarily disappear and re-emerge in response to external events, such as
demand spikes linked to regulatory deadlines or product launches.
An item removed due to low variance during a period of low incidence would
stop capturing the same factor when it returned, unless the qualitative
review process re-identifies it as emerging.
ADEMM, as formalized, is fundamentally reactive: it reacts to signals that
have already appeared in the data, but does not anticipate the return of
known factors that have temporarily left the instrument.
A future extension could treat items removed due to low variance as
"hibernated" rather than permanently excluded, allowing scheduled
reactivation during time windows associated with known seasonal patterns.

The actionability-based prioritization proposed in DP1, although useful for
keeping the instrument focused on factors the organization can intervene
on, has a relevant side effect: important structural factors that are out
of the organization's reach tend to be deprioritized or removed over the
iterations.
This occurred with the item related to waiting for external validation in
Iteration~5.
Organizations that wish to document systemic barriers of their clients,
even without the direct capacity to resolve them, should consider this
trade-off before adopting the method in its current form.

The work diary tested in this study showed low adherence in the format
evaluated, without automated reminders or participation incentives.
Future replications could test automated reminders, shorter formats
(one-sentence daily entries), or symbolic incentives before ruling out daily
self-report instruments as a complementary collection strategy.

Finally, the temporal traceability of the interviews relative to the
collection weeks was not preserved with sufficient precision to allow direct
quantitative cross-referencing between qualitative findings and specific
weekly percentages.
Future replications of ADEMM should explicitly record, for each interview,
which collection week it corresponds to or which set of cycles it spans, in
order to enable more precise integrated analyses between the two channels.

The main gap to be explored in future work, however, remains the absence of
a second instantiation of ADEMM in a different domain.
Without this, the three design principles should be treated as
\textit{nascent design theory}: the internal logic suggests broader
applicability, but this proposition still needs empirical evaluation in
other contexts.

\section{Threats to Validity}
\label{sec:threats-validity}

Following the framework of Wohlin et al.~\cite{wohlin2012experimentation},
this section discusses threats to construct, internal, external, and
reliability validity.
A final subsection addresses threats specific to the evaluation of ADEMM as
a DSR and ADR artifact, following the dimensions discussed by
Venable et al.~\cite{venable2012comprehensive}.

\subsection{Construct Validity}

The instrument measures self-reported perceptions of efficiency, not
objective productivity measures.
The results should be interpreted as factors perceived by participants as
influencing their efficiency, and not as direct measures of performance.
The interviews complemented the surveys by detailing the context of the
reported factors, but they also depend on the interpretation of both
participants and researchers.

The instrument was refined across five iterations.
Items added in later versions, such as those related to the use of
artificial intelligence, cannot be directly compared across the entire
twelve-week period and were interpreted only for the cycles in which they
were present.
Direct longitudinal comparisons were restricted to items that were
textually stable across consecutive iterations.

\subsection{Internal Validity}

The main threat to internal validity stems from the variation in
participation across collection weeks.
The number of respondents varied between 13 and 25 of the 27 participants,
which can introduce non-response bias: participants who were more
overloaded or more affected by certain barriers may have been less likely to
respond in specific weeks.
This risk is especially relevant in weeks with lower adherence, such as
Week~9, when some of the sharpest drops in the monitored factors also
occurred.

The study did not include a control group, a counterfactual condition, or a
static instrument for comparison.
The observed variations may be associated with changes in work demands, in
participants' context, in respondent composition, or in the instrument
itself, without it being possible to isolate any factor causally.

\subsection{External Validity}

The study was conducted with 27 participants from a single professional
training program in Brazil, linked to a single organization.
The results regarding the factors affecting efficiency cannot be
statistically generalized to other companies, countries, domains, or
software development contexts.

The investigated context has a particular characteristic: the organization
had no employment relationship with, or operational control over, the
participants, which differentiates the study from contexts with direct
control over processes, infrastructure, and working conditions.
ADEMM's design principles may not apply in the same way in those contexts.

The transferability of the method to other domains remains hypothetical.
The design principles should be interpreted as \textit{nascent design
theory}, whose applicability needs to be examined in future replications in
different contexts.

\subsection{Reliability}

The iterative adaptation of the instrument may have reduced comparability
across some cycles.
To mitigate this risk, factors were interpreted only for the periods in
which they were present in the instrument, and trend comparisons were
restricted to items stable across consecutive iterations.

The qualitative analysis combined deductive and inductive
strategies~\cite{braun2006thematic}, which made it possible to consider
previously identified factors while simultaneously recognizing emerging
patterns.
Even so, the interpretation of codes, categories, and themes may be
influenced by the researchers.
To increase traceability, records were kept of transcripts, assigned codes,
provisional categories, script versions, and instrument refinement
decisions.

\subsection{Threats Specific to DSR and ADR}

The categories of Wohlin et al.~\cite{wohlin2012experimentation} were
proposed for empirical studies in general and do not fully cover risks
specific to the evaluation of design artifacts.
This subsection considers five additional threats.

There is a risk of self-evaluation bias.
The same researchers who built ADEMM also participated in the formative and
summative evaluations, together with the person responsible for the
consulting practice.
There was no independent external evaluator.
Although this proximity is consistent with the logic of ADR, it reduces the
ability to rule out that the positive evaluation of the method was
influenced by the actors involved in its construction.

The co-evolution between artifact and organization makes it difficult to
separate the effects of ADEMM's design properties from the effects of the
maturing relationship between researchers and the \textit{problem owner}.
Since the instrument, the consulting practice's needs, and the
organizational decisions evolved across the same five iterations, it is not
possible to isolate with experimental precision whether the results stem
from the method or also from the gradual building of trust and
understanding between those involved.

The kernel theory for the first version of the instrument relied on
empirical generalizations from the literature on factors affecting
developer efficiency, treated as \textit{kernel theory} in a broad
sense~\cite{Sein2011}.
A more robust explanatory theory about the mechanisms by which these
factors influence efficiency could have produced an initial version with
greater explanatory power.

The design space was limited by access constraints to the participants'
employing companies.
The two alternatives actually evaluated were the work diary and tool
telemetry.
Other alternatives that would not require direct access to the
organizations, such as end-of-task forms or short daily logs, were not
evaluated.
ADEMM should not be interpreted as a solution that exhaustively explored all
available instrumentation alternatives.

Finally, the conclusion that iterative adaptation produced a more adequate
instrument than a static version was not directly tested against a
non-adaptive instrument.
This inference is supported by the documented decisions to retain, remove,
and include items and by the practical usefulness observed during the
study, but it remains a proposition to be examined in future replications
with comparative designs.
\section{Conclusion}
\label{sec:conclusion}

The literature on developer efficiency recognizes that the phenomenon is
multidimensional and dynamic~\cite{forsgren2021space,noda2023devex}, but
existing monitoring instruments respond to this dynamism in two
unsatisfactory ways: they are either cross-sectional and lose the temporal
evolution, or they are longitudinal and fixed and lose the dimensions that
emerge during the investigation itself.
This paper demonstrated that it is possible to combine longitudinality with
iterative adaptation of the instrument in a method applicable even when the
organization responsible for the follow-up has no access to the
participants' work environment.

ADEMM resolves this problem through three design decisions that DSR and ADR
leave open to the researcher: an operational item-removal criterion based on
two converging signals (low variance and decreasing usefulness as
recognized by the \textit{problem owner}), an inclusion criterion based on
recurring patterns in the qualitative channel, and an actionability-based
prioritization that guides which factors are worth monitoring given what
the organization can actually do with that information.
Formalized as DP1, DP2, and DP3, these principles are traceable to concrete
decisions documented throughout the study's five iterations, which
distinguishes them from generic recommendations with no empirical grounding.

The coexistence of these decisions within a single method is what
differentiates ADEMM from a generic application of DSR or ADR in the same
context.
The contribution table presented in Section~\ref{sec:ademm-contribution}
shows that each of the five points of distinction resolves a choice that
education consulting organizations would need to face with no applied
precedent if they started directly from Peffers et al. or Sein et al.
The resulting refinement was field-tested over four months, with 27
participants at distinct organizations, and produced sufficient evidence to
support concrete organizational decisions during its own execution, before
any publication.

Three specific research directions emerge directly from the documented
limitations.
The first is the replication of ADEMM in a domain other than software
development, a necessary condition for the three design principles to move
beyond \textit{nascent design theory} and consolidate as transferable design
knowledge.
The second is testing a pruning criterion that complements variance with a
severity indicator, in order to distinguish rare but severe factors from
genuinely irrelevant ones, a problem identified in DP3 during this study.
The third is comparing ADEMM with a fixed-instrument version applied to the
same context, to empirically examine whether iterative adaptation produces
different results than a static instrument would, an inference this study
supports logically but did not test in a controlled way.

Monitoring developer efficiency in real time, without access to the
repository, the pipeline, or the planning meeting of the team, requires a
method that relies more on the documented perception of the professionals
themselves than on metrics extracted from systems.
ADEMM shows that this reliance can be methodically structured: with
explicit review criteria, a qualitative channel that captures what closed
items do not anticipate, and a \textit{problem owner} who keeps the
instrument anchored to what the organization can actually do with the
information it collects.

\section*{Declaration of Use of Artificial Intelligence Tools}
The authors used Generative Artificial Intelligence tools (ChatGPT, by
OpenAI, and Claude, by Anthropic) to support the writing of this paper,
including generating textual excerpts from notes and data provided by the
authors, translating parts of the text, and linguistic and structural
review of the manuscript.
All content generated or reviewed with the aid of these tools was read,
verified, and edited by the authors, who take full responsibility for the
scientific accuracy, the integrity of the reported data, and the
conclusions presented in this work.



\bibliographystyle{IEEEtran}
\bibliography{references}

@inproceedings{dutra2015high,
  author    = {Dutra, Alessandra C. S. and Prikladnicki, Rafael and Fran{\c{c}}a, C{\'e}sar},
  title     = {What Do We Know about High Performance Teams in Software Engineering? {R}esults from a Systematic Literature Review},
  booktitle = {2015 41st Euromicro Conference on Software Engineering and Advanced Applications (SEAA)},
  year      = {2015},
  publisher = {IEEE},
  organization = {IEEE}
}

@book{reinertsen2009principles,
  author    = {Reinertsen, Donald G.},
  title     = {The Principles of Product Development Flow: Second Generation Lean Product Development},
  year      = {2009},
  publisher = {Celeritas Publishing}
}

@inproceedings{wagner2008systematic,
  author    = {Wagner, Stefan and Ruhe, Melanie},
  title     = {A Systematic Review of Productivity Factors in Software Development},
  booktitle = {Proceedings of the 2nd International Workshop on Software Productivity Analysis and Cost Estimation (SPACE 2008)},
  year      = {2008}
}

@article{forsgren2021space,
  author    = {Forsgren, Nicole and Storey, Margaret-Anne and Maddila, Chandra
               and Zimmermann, Thomas and Houck, Brian and Butler, Jenna},
  title     = {The {SPACE} of Developer Productivity: There's More to It Than You Think},
  journal   = {{ACM} Queue},
  volume    = {19},
  number    = {1},
  pages     = {20--48},
  year      = {2021},
  doi       = {10.1145/3454122.3454124}
}

@article{greiler2023devex,
  author    = {Greiler, Michaela and Storey, Margaret-Anne and Noda, Abi},
  title     = {An Actionable Framework for Understanding and Improving Developer Experience},
  journal   = {{IEEE} Transactions on Software Engineering},
  volume    = {49},
  number    = {4},
  pages     = {1411--1425},
  year      = {2023},
  doi       = {10.1109/TSE.2022.3175660}
}

@article{kuutila2021individual,
  author    = {Kuutila, Miikka and M{\"a}ntyl{\"a}, Mika V. and Claes, Ma{\"e}lick
               and Elovainio, Marko and Adams, Bram},
  title     = {Individual Differences Limit Predicting Well-being and Productivity
               Using Software Repositories: A Longitudinal Industrial Study},
  journal   = {Empirical Software Engineering},
  volume    = {26},
  number    = {5},
  pages     = {1--38},
  year      = {2021},
  doi       = {10.1007/s10664-021-09977-1}
}

@article{russo2024longitudinal,
  author    = {Russo, Daniel and Hanel, Paul H. P. and van Berkel, Niels},
  title     = {Understanding Developers Well-Being and Productivity:
               A 2-year Longitudinal Analysis during the {COVID-19} Pandemic},
  journal   = {{ACM} Transactions on Software Engineering and Methodology},
  volume    = {33},
  number    = {3},
  year      = {2024},
  doi       = {10.1145/3638244}
}

@book{sadowski2019rethinking,
  editor    = {Sadowski, Caitlin and Zimmermann, Thomas},
  title     = {Rethinking Productivity in Software Engineering},
  publisher = {Apress},
  year      = {2019},
  doi       = {10.1007/978-1-4842-4221-6}
}

@article{noda2023devex,
  author    = {Noda, Abi and Storey, Margaret-Anne and Forsgren, Nicole and Greiler, Michaela},
  title     = {{DevEx}: What Actually Drives Productivity},
  journal   = {{ACM} Queue},
  volume    = {21},
  number    = {2},
  pages     = {35--53},
  year      = {2023},
  doi       = {10.1145/3595878}
}

@book{demarco1987peopleware,
  author    = {DeMarco, Tom and Lister, Timothy},
  title     = {Peopleware: Productive Projects and Teams},
  publisher = {Dorset House},
  year      = {1987},
  address   = {New York, NY}
}

@book{forsgren2018accelerate,
  author    = {Forsgren, Nicole and Humble, Jez and Kim, Gene},
  title     = {Accelerate: The Science of Lean Software and DevOps},
  publisher = {IT Revolution Press},
  year      = {2018},
  address   = {Portland, OR}
}

@article{meyer2019,
  author  = {Meyer, Andr{\'e} N. and Barr, Elaine T. and Fritz, Thomas and Zimmermann, Thomas},
  title   = {Today Was a Good Day: The Daily Life of Software Developers},
  journal = {IEEE Transactions on Software Engineering},
  year    = {2019},
  volume  = {47},
  number  = {5},
  pages   = {3505--3522},
  note    = {TODO: verify volume, issue and page numbers}
}

@article{kalliamvakou2022,
  author  = {Kalliamvakou, Eirini},
  title   = {Research: Quantifying {GitHub Copilot}'s Impact on Developer
             Productivity and Happiness},
  journal = {GitHub Blog},
  year    = {2022},
  note    = {TODO: replace with peer-reviewed citation if available.
             URL: https://github.blog/2022-09-07-research-quantifying-github-copilots-impact-on-developer-productivity-and-happiness/}
}

@article{meyer2014software,
  title={Software developers' perceptions of productivity},
  author={Meyer, Andr{\'e} N and Fritz, Thomas and Murphy, Gail C and Zimmermann, Thomas},
  booktitle={Proceedings of the 22nd ACM SIGSOFT international symposium on foundations of software engineering},
  pages={19--29},
  year={2014}
}

@article{coelho2025software,
  title={Software Developers’ Perceptions of Productivity: An Industry-focused Study},
  author={Coelho, Murilo and Reinbold, Isabelle and Sancho, Lizie and Paixao, Matheus and Ara{\'u}jo, Allysson Allex and Freire, S{\'a}vio},
  booktitle={Simp{\'o}sio Brasileiro de Qualidade de Software (SBQS)},
  pages={12--22},
  year={2025},
  organization={SBC}
}

@inproceedings{cheng2022improves,
  title={What improves developer productivity at google? code quality},
  author={Cheng, Lan and Murphy-Hill, Emerson and Canning, Mark and Jaspan, Ciera and Green, Collin and Knight, Andrea and Zhang, Nan and Kammer, Elizabeth},
  booktitle={Proceedings of the 30th ACM Joint European Software Engineering Conference and Symposium on the Foundations of Software Engineering},
  pages={1302--1313},
  year={2022}
}

@article{razzaq2024systematic,
  title={A systematic literature review on the influence of enhanced developer experience on developers' productivity: Factors, practices, and recommendations},
  author={Razzaq, Abdul and Buckley, Jim and Lai, Qin and Yu, Tingting and Botterweck, Goetz},
  journal={ACM Computing Surveys},
  volume={57},
  number={1},
  pages={1--46},
  year={2024},
  publisher={ACM New York, NY}
}

@article{d2024measuring,
  title={Measuring developer experience with a longitudinal survey},
  author={D’Angelo, Sarah and Lin, Jessica and Dicker, Jill and Egelman, Carolyn and Hodges, Maggie and Green, Collin and Jaspan, Ciera},
  journal={IEEE software},
  volume={41},
  number={4},
  pages={19--24},
  year={2024},
  publisher={IEEE}
}

@inproceedings{canedo2019factors,
  title={Factors affecting software development productivity: An empirical study},
  author={Canedo, Edna Dias and Santos, Giovanni Almeida},
  booktitle={Proceedings of the XXXIII Brazilian Symposium on Software Engineering},
  pages={307--316},
  year={2019}
}

@inproceedings{coutinho2024role,
  title={The role of generative ai in software development productivity: A pilot case study},
  author={Coutinho, Mariana and Marques, Lorena and Santos, Anderson and Dahia, Marcio and Fran{\c{c}}a, Cesar and de Souza Santos, Ronnie},
  booktitle={Proceedings of the 1st ACM International Conference on AI-Powered Software},
  pages={131--138},
  year={2024}
}

@book{wohlin2012experimentation,
  title={Experimentation in software engineering},
  author={Wohlin, Claes and Runeson, Per and H{\"o}st, Martin and Ohlsson, Magnus C and Regnell, Bj{\"o}rn and Wessl{\'e}n, Anders and others},
  volume={236},
  year={2012},
  publisher={Springer}
}

@article{hevner2004design,
  author  = {Hevner, Alan R. and March, Salvatore T. and Park, Jinsoo and Ram, Sudha},
  title   = {Design Science in Information Systems Research},
  journal = {MIS Quarterly},
  volume  = {28},
  number  = {1},
  pages   = {75--105},
  year    = {2004},
  doi     = {10.2307/25148625}
}

@article{braun2006thematic,
  author  = {Braun, Virginia and Clarke, Victoria},
  title   = {Using Thematic Analysis in Psychology},
  journal = {Qualitative Research in Psychology},
  volume  = {3},
  number  = {2},
  pages   = {77--101},
  year    = {2006},
  doi     = {10.1191/1478088706qp063oa}
}

@article{braun2006using,
  title={Using thematic analysis in psychology},
  author={Braun, Virginia and Clarke, Victoria},
  journal={Qualitative research in psychology},
  volume={3},
  number={2},
  pages={77--101},
  year={2006},
  publisher={Taylor \& Francis}
}

@article{Sein2011,
  title={Action design research},
  author={Sein, Maung K and Henfridsson, Ola and Purao, Sandeep and Rossi, Matti and Lindgren, Rikard},
  journal={MIS Q},
  volume={35},
  number={4},
  pages={1099--1099},
  year={2011}
}

@book{Wieringa2014,
  title={Design science methodology for information systems and software engineering},
  author={Wieringa, Roel},
  year={2014},
  publisher={Springer}
}

@article{gregor2020research,
  title={Research perspectives: the anatomy of a design principle},
  author={Gregor, Shirley and Chandra Kruse, Leona and Seidel, Stefan},
  journal={Journal of the Association for Information Systems},
  volume={21},
  number={6},
  pages={2},
  year={2020}
}

@article{bolger2003diary,
  title={Diary methods: Capturing life as it is lived},
  author={Bolger, Niall and Davis, Angelina and Rafaeli, Eshkol},
  journal={Annual review of psychology},
  volume={54},
  number={1},
  pages={579--616},
  year={2003},
  publisher={Annual Reviews 4139 El Camino Way, PO Box 10139, Palo Alto, CA 94303-0139, USA}
}

@article{galesic2009effects,
  title={Effects of questionnaire length on participation and indicators of response quality in a web survey},
  author={Galesic, Mirta and Bosnjak, Michael},
  journal={Public opinion quarterly},
  volume={73},
  number={2},
  pages={349--360},
  year={2009},
  publisher={Oxford University Press}
}

@article{peffers2007design,
  title={A design science research methodology for information systems research},
  author={Peffers, Ken and Tuunanen, Tuure and Rothenberger, Marcus A and Chatterjee, Samir},
  journal={Journal of management information systems},
  volume={24},
  number={3},
  pages={45--77},
  year={2007},
  publisher={Taylor \& Francis}
}

@article{lynn2009methods,
  title={Methods for longitudinal surveys},
  author={Lynn, Peter},
  journal={Methodology of longitudinal surveys},
  pages={1--19},
  year={2009},
  publisher={Wiley Online Library}
}

@inproceedings{venable2012comprehensive,
  title={A comprehensive framework for evaluation in design science research},
  author={Venable, John and Pries-Heje, Jan and Baskerville, Richard},
  booktitle={International conference on design science research in information systems},
  pages={423--438},
  year={2012},
  organization={Springer}
}

\end{document}